\documentstyle[12pt,preprint,aps,floats]{revtex}

\tighten

\begin{document}

\preprint{
\noindent
\hfill
\begin{minipage}[t]{3in}
\begin{flushright}
LBNL--39578\\
UCB--PTH--96/50\\
SCIPP--96/72\\
RU--96--97\\
hep-ph/9612397\\
\vspace*{2cm}
\end{flushright}
\end{minipage}
}

\draft

\title{A  wide scalar neutrino resonance and $\bbox{b\bar{b}}$ 
production at LEP}
\author{Jens Erler$^{a}$, Jonathan L. Feng$^{b}$, and Nir 
Polonsky$^{c}$}
\address{${}^{a}$Santa Cruz Institute for Particle Physics, 
University of California, Santa Cruz, CA 95064}
\address{${}^{b}$Theoretical Physics Group, E.O. Lawrence Berkeley
National Laboratory \\ and Department of Physics, University of
California, Berkeley, CA 94720}
\address{${}^{c}$Department of Physics and Astronomy, 
Rutgers University, Piscataway, NJ 08855-0849 
} 
\date{December 1996}

\maketitle

\begin{abstract}
In supersymmetric models with $R$-parity violation, scalar neutrinos
$\tilde{\nu}$ may be produced as $s$-channel resonances in $e^+e^-$
colliders. We note that within current constraints, the scalar neutrino
may have a width of several GeV into $b\bar{b}$ and be produced with
large cross section, leading to a novel supersymmetry discovery signal
at LEP II.  In addition, if $m_{\tilde{\nu}} \approx m_Z$, such a
resonance necessarily increases $R_b$ and reduces $A_{FB}(b)$,
significantly improving the fit to electroweak data. Bounds from $B$
meson and top quark decays are leading constraints, and we stress the
importance of future measurements.
\end{abstract}
 


One of the important goals of future collider experiments is to search
for and possibly discover supersymmetry.  In the most widely analyzed
supersymmetric extension of the standard model (SM), the superpotential
is assumed to be $W = h_E H_1 LE^c + h_D H_1 QD^c - h_U H_2 Q U^c - \mu
H_1 H_2$, where the lepton and quark superfields $L = (N, E)$, $E^c$, $Q
= (U, D)$, $U^c$, and $D^c$ contain the SM fermions $f$ and their scalar
partners $\tilde{f}$, and generation indices have been omitted.  This
superpotential conserves $R$-parity, $R_P = (-1)^{2J + 3B + L}$, where
$J$, $B$, and $L$ are spin, baryon number and lepton number,
respectively.  $R_P$ conservation strongly restricts the phenomenology,
as it implies that superpartners must be produced in pairs and that the
lightest supersymmetric particle (LSP) is stable.

The superpotential above, however, is not the most general allowed by
gauge invariance and renormalizability.  In particular, as the
superfields $H_1$ and $L$ have the same quantum numbers, the
$R_P$-violating $(\rlap{$\not$}R_P )$ terms

\begin{equation}
W_{\not L} = \lambda L L E^c + \lambda' L Q D^c 
\label{superpotential}
\end{equation}
are allowed. We will consider these couplings, the most general
trilinear $\rlap{$\not$}R_P$ terms that violate lepton number but not
baryon number. (Note that proton stability requires only approximate
conservation of either lepton or baryon number.)  Such terms have a
number of interesting properties, including the possibility of providing
new avenues for neutrino mass generation\cite{hs,np}, which otherwise
must be attributed to some grand-scale sector of the theory, as in, {\em
e.g.}, see-saw models. Here, we focus on another implication of these
terms, namely, the possibility of sneutrino resonances at $e^+e^-$
colliders\cite{dh88,bgh}.  Such resonances offer the unique opportunity
to probe supersymmetric particle masses up to $\sqrt{s}$, which, at LEP
II, is well into the range typically predicted for slepton masses.

The superpotential of Eq.~(\ref{superpotential}) generates couplings

\begin{eqnarray}
{\cal L}_{\not L} 
&=&\lambda_{ijk} \left[ \tilde{e}_L^i \overline{e_R^k} \nu_L^j
   + \tilde{\nu}_L^j \overline{e_R^k} e_L^i
+ (\tilde{e}_R^k)^\ast \overline{(e_L^i)^c} \nu_L^j \right]  \nonumber \\
&-&\lambda_{ijk} \left[ \tilde{\nu}_L^i \overline{e_R^k} e_L^j
         + \tilde{e}_L^j \overline{e_R^k} \nu_L^i
         + (\tilde{e}_R^k)^\ast \overline{(\nu_L^i)^c} e_L^j \right] 
\nonumber \\
&+& \lambda'_{lmn} {V^{\ast pm}_{CKM}} 
\left[ \tilde{e}_L^l \overline{d_R^n} u_L^p
  + \tilde{u}_L^p \overline{d_R^n} e_L^l
  + (\tilde{d}_R^n)^\ast \overline{(e_L^l)^c} u_L^p \right] \nonumber \\
&-&\lambda'_{lmn} \left[ \tilde{\nu}_L^l \overline{d_R^n} d_L^m
  + \tilde{d}_L^m \overline{d_R^n} \nu_L^l
  + (\tilde{d}_R^n)^\ast \overline{(\nu_L^l)^c} d_L^m \right]
\label{lagrangian} 
\ , 
\end{eqnarray}
where $V_{CKM}$ is the Cabibbo-Kobayashi-Maskawa matrix, we assume
$f$--$\tilde{f}$ alignment, $i<j$, and the other generation indices are
arbitrary.  As we will concentrate on the sneutrino $\tilde{\nu}$, we
have chosen the basis in which $NDD^c$ is diagonal; implications of
choosing another basis will be discussed below.

The interactions of Eq.~(\ref{lagrangian}) imply that sneutrinos may be
produced as $s$-channel resonances with cross section

\begin{equation}
\sigma(e^{+}e^{-} \rightarrow \tilde{\nu} \rightarrow X)
= \frac{8\pi s}{m_{\tilde{\nu}}^{2}} \frac
{\Gamma_{\tilde{\nu}\rightarrow e^+e^-}
\Gamma_{\tilde{\nu}\rightarrow X}}
{(s - m_{\tilde{\nu}}^{2})^{2} +
m_{\tilde{\nu}}^{2}\Gamma_{\tilde{\nu}}^{2}} \ .
\label{xsection}
\end{equation}
(Lepton pair production may also receive contributions from $t$-channel
$\tilde{\nu}$ exchange.)  If the sneutrino is the LSP, it decays to
pairs of charged leptons or down-type quarks with width

\begin{equation}
\Gamma_{\tilde{\nu}\rightarrow f\bar{f}'}
= N_{c} \frac{g^{2}}{16\pi}m_{\tilde{\nu}} \ ,
\label{width1}
\end{equation}
where $N_c$ is the color factor and $g$ is the relevant
$\rlap{$\not$}R_P$ coupling.  Decays to neutrinos and up-type quarks are
prohibited by gauge invariance.  On the other hand, if the LSP is the
lightest neutralino $\chi^{0}$, the sneutrino may also decay through
$\tilde{\nu} \rightarrow \nu \chi^{0}$, with partial width $\sim 0.1 -
1$ GeV for $m_{\tilde{\nu}} \sim 100 - 200$ GeV ~\cite{bgh}.  The
neutralino $\chi^0$ then decays to three SM fermions through
$\rlap{$\not$}R_P$ interactions.

In this study, motivated by the Yukawa renormalization of the scalar
spectrum, which typically leaves the third generation scalar fields
lighter than the first two, we focus on the possibility of a
$\tilde{\nu}_{\tau}$ resonance.  In addition, we concentrate on
$\tilde{\nu}_{\tau}$ decays to $b\bar{b}$ pairs, as the possibility of a
wide resonance will be evident from considerations of this channel
alone.  We therefore consider the scenario in which the non-zero
couplings of Eq.~(\ref{lagrangian}) are $\lambda_{131}$ and
$\lambda'_{333}$, and, for simplicity, we take these to be real.  Note,
however, that a $\tilde{\nu}_{\mu}$ resonance and decays to other final
states, {\em e.g.}, $b\bar{d}$ and $b\bar{s}$, though more highly
constrained, are also possible in principle.

Bounds on $\lambda_{131}$ and $\lambda'_{333}$, taken individually, have
been considered previously, and the strongest of these are
$\lambda_{131} < 0.10 \, [m_{\tilde{e}_R}/100 \text{ GeV}]$ from
$\Gamma(\tau\to e \nu \bar{\nu}) / \Gamma(\tau\to \mu \nu
\bar{\nu})$\cite{bgh}, and $\lambda'_{333} < 0.6 - 1.3$ (2$\sigma$) from
$R_{\tau}$\cite{bes}, where the range is for $m_{\tilde{q}} =
2m_{\tilde{l}} = 300$ GeV -- 1 TeV.  These and the new bounds derived
below are collected in Table~\ref{table:t1}.

\begin{table}
\caption{Upper bounds on the couplings $\lambda_{131}$ and 
$\lambda'_{333}$.}
\begin{tabular}{lll}
Coupling & Upper Bound & Process \\
\hline
$\lambda_{131}$ & 0.10 $[m_{\tilde{e}_R}/100 \text{ GeV}]$ & 
$\frac{\Gamma(\tau\to e \nu \bar{\nu})}{\Gamma(\tau\to \mu \nu
\bar{\nu})}$\cite{bgh} 
\\
$\lambda'_{333}$ & $0.6 - 1.3$ ($2\sigma$) & 
$R_{\tau}$ ($m_{\tilde{q}} = 0.3-1$ TeV) \cite{bes}
\\
$\lambda'_{333}$ & $0.96 \left[m_{\tilde{b}_R}/ 300 \text{ GeV}\right] 
$ & $ B \to \tau \bar{\nu} X$
\\
$\lambda_{131}\lambda'_{333}$ & 0.075 $\left[{m_{\tilde{\tau}_L}/ 100
\text{ GeV}}\right]^2$ & $B^- \to e \bar{\nu}$
\\
\end{tabular}
\label{table:t1}
\end{table}

In the limit of large scalar masses, the interactions of
Eq.~(\ref{lagrangian}) induce many four-fermion operators, some of which
mediate meson decays.  A competitive bound on $\lambda'_{333}$ arises
{}from $B \to \tau \bar{\nu} X$ through the $\rlap{$\not$}R_P$ operator
$ - \frac{\lambda'^{2}_{333}}{m_{\tilde{b}_R}^2}V_{cb}
(\overline{({\nu_{\tau}}_L)^c} b_L)(\overline{c_L} (\tau_L)^c)$.  After
a Fierz transformation, this is seen to interfere constructively with
the SM operator to give

\begin{equation}
-V_{cb} \left[\frac{4 G_F}{\sqrt{2}} +
\frac{\lambda'^{2}_{333}}{2 m^2_{\tilde{b}_{R}}}\right]
\bar{c}_L\gamma_{\mu} b_L \bar{\tau}_L \gamma^{\mu} \nu_{\tau} \ .
\end{equation}
The experimental bound and SM prediction for $B(B\to \tau \bar{\nu}X)$
are $2.68 \pm 0.34\%$ \cite{Btauexp} and $2.30 \pm 0.25\%$\cite{Btauth},
respectively. Simply combining these errors in quadrature and demanding
that the $\rlap{$\not$}R_P$-enhanced rate be below the current upper
bound, we find the constraint $\lambda'_{333} < 0.96 \,
[m_{\tilde{b}_R}/ 300 \text{ GeV}]$.

Meson decays also bound the product $\lambda_{131} \lambda'_{333}$,
which enters in the cross section of Eq.~(\ref{xsection}).  The operator
$- \frac{\lambda_{131}\lambda'_{333}} {m_{\tilde{\tau}_L}^2} V_{pb}
(\overline{e_R} \nu_L)(\overline{u_L^p} b_R)$ is most stringently
bounded by taking $p=u$ and considering $B^- \to e^- \bar{\nu}$.  The SM
contribution to this decay is helicity-suppressed and negligible.  The
$\rlap{$\not$}R_P$ decay width may be calculated using $\langle 0 |
\bar{u}\gamma^5 b | B^- \rangle = -i f_B m_B^2/m_b$\cite{hou} to be

\begin{equation}
\Gamma = \frac{1}{64\pi} |V_{ub}|^2 \lambda^2_{131}\lambda'^{2}_{333}
\frac{1}{ {m_{\tilde{\tau}_L}^4}} \frac{f_B^2 m_B^5}{m_b^2} \ .
\end{equation}
Applying the current bound $B(B^- \to e^- \bar{\nu}) < 1.5\times
10^{-5}$\cite{PDG}, and taking $V_{ub} > 0.0024$, $f_B > 140$ MeV, and
$m_b = 4.5$ GeV, we find the upper bound $\lambda_{131}\lambda'_{333}<
0.075 \, [{{m_{\tilde{\tau}_L}}}/100 \text{ GeV}]^2$.

The $\rlap{$\not$}R_P$ couplings are also constrained by other $B$
decays, $\Upsilon$ decays, and the collider bound on $m_{\nu_{\tau}}$.
These bounds, however, are not competitive with those discussed above.
In addition, under the assumption that only $\lambda'_{333}$ is
non-zero, there are no contributions to $K^0 - \bar{K}^0$ and $B^0 -
\bar{B}^0$ mixing, and $D^0 - \bar{D}^0$ mixing gives an extremely weak
constraint.  If we had worked in the basis in which $EUD^c$ is diagonal
and considered the possibility of only $\lambda'^{U}_{333}$ non-zero,
the $\rlap{$\not$}R_P$ interactions would contribute to neutral $K$ and
$B$ meson mixing, as well as to $\Delta B = 1$ neutral current
decays. The bound from $B^0 - \bar{B}^0$ mixing is stronger than the one
{}from $K^0 - \bar{K}^0$ mixing, and is $\lambda'^{U}_{333} < 1.1$ for
scalar masses of 100 GeV\cite{ag}.  $B(B^0 \to e^+e^-) < 5.9 \times
10^{-6}$\cite{PDG} and $B(B^0 \to K^0 e^+e^-) < 3\times
10^{-4}$\cite{PDG} both imply $\lambda_{131}\lambda'^{U}_{333} \alt 0.03
\, [{m_{\tilde{\nu}_{\tau}}}/100 \text{ GeV}]^{2}$.  We see then that
numerically the bounds are fairly basis independent.

An independent set of constraints arises from the exotic top quark decay
mode $t_{L} \rightarrow b_R \tilde{\tau}_{L}^{+}$, assuming it is
kinematically allowed.  (Note that SU(2) invariance requires
$m_{\tilde{\tau}_{L}} \simeq m_{\tilde{\nu}_{\tau}}$.)  For $m_t = 175$
GeV,

\begin{equation}
R_t \equiv \frac
{\Gamma_{t \rightarrow b\tilde{\tau}}}
{\Gamma_{t \rightarrow bW}}=
1.12 \, \lambda'^2_{333}\left[
1 - \left(\frac{m_{\tilde{\tau}_L}}{175\text{ GeV}}\right)^2
\right]^{2}.
\label{rt}
\end{equation}
If the sneutrino is the LSP, the three-body decays $\tilde{\tau} \to
\tilde{\nu}_{\tau} f\bar{f}'$ and $\tilde{\tau} \to Wb\bar{b}$ are
sufficiently phase space-suppressed that the dominant decay mode is
either $\tilde{\tau} \to \bar{c}b$ or $\tilde{\tau} \to e \bar{\nu}_e$.
The former is suppressed by $|V_{cb}|^2 \lambda'^2_{333}$, and the
latter by $\lambda^2_{131}$.  As top constraints will be important only
for large $\lambda'_{333}$, we first assume that the $\bar{c}b$ mode is
dominant.  This new decay mode alters the number of $t\bar{t}$ events
expected in each channel, both through an enhancement of the percentage
of hadronic decays and through the increased probability of $b$-tagging
events with $b$-rich $\tilde{\tau}$ decays.  For each channel, we denote
the number of events expected in the presence of $\tilde{\tau}$ decays
relative to the number expected in the SM as
\begin{equation}
R_B (x) \equiv \frac
{B(t\bar{t} \rightarrow X; x)}{B(t\bar{t} \rightarrow X ;x = 0)} \ ,
\label{counting}
\end{equation}
where $x = B(t \rightarrow b\tilde{\tau}) = R_{t}/(1 + R_{t})$.  The
expressions for these ratios are given in Table \ref{table:t2}, where
$\varepsilon_{m,n}$ denotes the probability of tagging at least $m$ of
$n$ $b$ jets.  The SVX $b$-tagging efficiency for $t\bar{t}$ events is
$\varepsilon_{1,2} = 41 \pm 4\%$~\cite{top}.  Crudely neglecting the
dependence of $b$-tagging efficiencies on the number of jets and jet
momenta, the remaining $b$-tagging efficiencies are then determined by
$\varepsilon_{1,2}$ to be, {\em e.g.}, $\varepsilon_{1,3} \approx 55\%$
and $\varepsilon_{1,4} \approx 65\%$. This approximation is conservative
when $m_{\tilde{\tau}}$ approaches $m_t$, as the soft $b$ jets lower
$\varepsilon_{1,n}$, but we will ignore this effect here.

\begin{table}
\caption{The ratios $R_B(x)$, where $H$ is the $W$ hadronic branching 
fraction, and $\varepsilon_{m,n}$ are $b$-tagging efficiencies.}
\begin{tabular}{ll}
Channel & $R_B(x)$ \\ \hline
dilepton & $ (1 -x)^2$ \\
lepton $+$ jets &
$(1 - x)^2 + \frac{\varepsilon_{1,3}}{\varepsilon_{1,2}} 
\frac{1}{H} x(1-x)$ \\
all-hadronic &
$(1 - x)^2 + \frac{\varepsilon_{1,3}}{\varepsilon_{1,2}} 
\frac{1}{H} 2x(1-x)
+\frac{\varepsilon_{1,4}}{\varepsilon_{1,2}} \frac{1}{H^2}x^2 $ \
\end{tabular}
\label{table:t2}
\end{table}

Based on an event sample of 110 pb$^{-1}$, the production cross section
has been measured by CDF to be $\sigma[t\bar{t}]_{\mbox{\tiny exp}} =
8.3^{+4.3}_{-3.3}$, $6.4^{+2.2}_{-1.8}$, and $10.7^{+7.6}_{-4.0}$ pb in
the dilepton, SVX lepton + jets, and all-hadronic channels, respectively
~\cite{top}. The SM theoretical expectation for $m_t = 175$ GeV is
$\sigma[t\bar{t}]_{\mbox{\tiny QCD}} = 5.5^{+0.1}_{-0.4}$
pb~\cite{qcd}. The coupling $\lambda'_{333}$ may then be bounded by
requiring that $R_B(x)$ lie within the measured range of
$\sigma[t\bar{t}]_{\mbox{\tiny exp}} / \sigma[t\bar{t}]_{\mbox{\tiny
QCD}}$ for each channel.  The 2$\sigma$ upper bound from the dilepton
channel is $\lambda'_{333} < 1.3$ for $m_{\tilde{\tau_{L}}} = 100$ GeV
and is degraded to $\lambda'_{333} < 3.2$ for $m_{\tilde{\tau_{L}}} =
150$ GeV. Significantly weaker constraints arise from the other
channels.  We see that these counting experiments currently give weak
constraints.  In addition, given the number of $t\bar{t}$ candidate
events at present and the low probability of tagging 3 or more $b$ jets,
there are no available limits from such multi-$b$-tagged events;
eventually, these limits may strongly constrain
$\lambda'_{333}$~\cite{ag}.

A more promising approach is to examine kinematic parameters in
$t\bar{t}$ events, {\em e.g.}, the reconstructed $W$ mass in lepton +
jets events with a second loosely tagged $b$~\cite{top}. (The two
untagged jets define $m_W$.)  In a sample of $N$ such events, an upper
bound of 3 events outside the $m_W$ peak would imply $\varepsilon_{2,3}/
\varepsilon_{2,2} [R_t /H] < 3/N$, where we have ignored differences 
between the usual and loose $b$-tag efficiencies.  Currently, with just
10 events, this gives $\lambda'_{333}\alt 0.4 \, (1.0)$ for
$m_{\tilde{\tau}_{L}} = 100 \, (150)$ GeV.  Such kinematic analyses may
therefore provide strong constraints on $\rlap{$\not$}R_P$ couplings in
the future.

If $\lambda_{131}$ is large enough that the decay mode $\tilde{\tau} \to
e\bar{\nu}_e$ dominates, the $\rlap{$\not$}R_P$ decay violates $e$-$\mu$
universality in $t$ decays\cite{ag}, leading to the constraint
$\lambda'_{333} < 0.35 \, (0.90)$ for $m_{\tilde{\tau}_{L}} = 100 \,
(150)$ GeV.  Finally, we briefly comment on the neutralino LSP case,
which was discussed in Ref.~\cite{ag}.  In our case, the decay
$\tilde{\tau} \to \tau \chi^0 \to \tau\nu bb$ may be constrained
by counting experiments as above, but with the substitutions
$\varepsilon_{1,3} \to \varepsilon_{1,4}$ and $\varepsilon_{1,4} \to
\varepsilon_{1,6}$ in $R_B(x)$.  The dilepton channel again gives the
strongest bounds, and, as these are independent of $b$-tagging
efficiencies, we again find $\lambda'_{333} < 1.3 \, (3.2)$ for
$m_{\tilde{\tau}_{L}} = 100 \, (150)$ GeV.  With more $t\bar{t}$
candidate events, one could also constrain the $\tau$ excess in the
different channels.

We conclude therefore that couplings $\lambda_{131} \sim {\cal O}(0.1)$
and $\lambda'_{333} \sim {\cal O}(1)$ are consistent with all of the
constraints considered above.  With such couplings, a sneutrino with
$m_{\tilde{\nu}} \alt 190$ GeV will be singly produced at LEP II and may
be observed as a resonance with $\Gamma_{\tilde{\nu}} \approx 6.0 \text{
GeV} \times \lambda'^{2}_{333}\, [m_{\tilde{\nu}}/100 \text{ GeV}]$.
(Of course, if light enough, sneutrinos may also be pair-produced at LEP
II and will be easily discovered through their $b$ quark
signature\cite{Barger}.)  Such a resonance could be discovered as a peak
in the initial state radiation (ISR)-induced tails of
$\sqrt{s_{\text{eff}}}$ distributions in two jet events. For fixed
values of $m_{\tilde{\nu}}$, $\lambda_{131}$, and $\lambda'_{333}$, one
may estimate the effective luminosity at $\sqrt{s_{\text{eff}}}
\approx m_{\tilde{\nu}}$ required to discover such a peak. We demand a 
5$\sigma$ $b\bar{b}$ excess in a bin centered at $m_{\tilde{\nu}}$ of
width $\max \{2 \text{ GeV}, 2\Gamma_{\tilde{\nu}}\}$ and assume a
tagging efficiency of 40\% for $b\bar{b}$ events.  The required
luminosities have non-trivial dependences on the various parameters; a
sample of results is given in Table~\ref{table:tscan}.  Examples $D$ and
$F$ are already probed, in principle, by the luminosity available in the
ISR tails of the $\sqrt{s} = 130-174$ GeV runs, provided that the data
of the four experiments are combined.  A total integrated luminosity of
$\sim 500 \text{ pb}^{-1}$ per detector at $\sqrt{s} \approx 190$ GeV
will yield an effective luminosity sufficient to probe examples $A$ and
$C$.  Coverage of the parameter space may be further improved by runs at
lower beam energies with a luminosity of ${\cal{O}}(10 {\text{
pb}}^{-1})$, which would probe examples $B$ and $E$.  Such runs would
also probe the parameter space if the neutralino is the LSP and
$\tilde{\nu} \to \nu \chi^{0} \to\nu\nu bb$, as the final state in this
case is still characterized by an excess of $b$ jets, but with a smeared
$\sqrt{s_{\text{eff}}}$ spectrum.  We conclude that values of
$\lambda_{131}\lambda'_{333}$ more than two orders of magnitude below
current bounds could be probed by the LEP experiments.  These analyses
and searches therefore offer exciting, if unconventional, possibilities
for the discovery of supersymmetry.
\begin{table}
\caption{The effective integrated luminosity ${\cal L}_{\text{eff}}$
per 1 GeV bin required to discover a sneutrino resonance with
$m_{\tilde{\nu}}$, $\lambda_{131}$, and $\lambda'_{333}$ as given.}
\begin{tabular}{lcccccc}
& $A$ & $B$ & $C$ & $D$ & $E$ & $F$ \\ \hline
$m_{\tilde{\nu}}$ (GeV)& 110  & 110  & 110  & 145  & 145  & 145  \\
$\lambda_{131}$        & 0.01 & 0.005 & 0.005 & 0.01 & 0.005 & 0.005 \\
$\lambda'_{333}$       & 1.0  & 1.0  & 0.1  & 1.0  & 1.0  & 0.1  \\
${\cal L}_{\text{eff}}$ (pb$^{-1}$/\text{GeV}) 
		       & 1.1 & 18 & 1.3 &  0.45  & 7.1  & 0.43 
\end{tabular}
\label{table:tscan}
\end{table}
 
Finally, a most interesting window for the sneutrino resonance exists
near the $Z$ pole, illustrating the possibility of new physics hidden by
the $Z$ resonance~\cite{jens}. It is intriguing that a $\tilde{\nu}$
resonance in this window necessarily increases $R_b$ and decreases
$A_{FB}(b)$, in accord with current measurements\cite{ICHEP}. In
addition, gauge invariance prohibits the $\tilde{\nu}$ from directly
affecting $c\bar{c}$ production, and we have explicitly confirmed that
direct effects on leptonic observables, {\em e.g.}, the $t$-channel
$\tilde{\nu}$ contribution to $A_{FB}(e)$, are negligible given the
constraints on $\lambda_{131}$.  We have performed a $Z$ lineshape fit
including the sneutrino resonance in the sneutrino LSP scenario.  Our
treatment and approximations follow closely those described in
Ref.~\cite{jens}, where four-fermion operators with cross sections
depending linearly on $s$ were studied; here, we superimpose a second
$s$ resonance on that of the $Z$.

We restricted our lineshape fit to the published data of only one LEP
group (L3) \cite{L3} from the years 1990 -- 1992. This will suffice, as
here we are interested in the changes caused by the introduction of a
sneutrino with $m_{\tilde{\nu}} \approx m_Z$. We also included the SLD
Collaboration's determination of the left-right polarization asymmetry,
$A_{LR}$, recorded during 1992 -- 1995\cite{SLD}, and results of the LEP
and SLD Heavy Flavor Groups, as reported in Ref.~\cite{ICHEP}.  The
latter include the $Z \to b\bar{b}$ branching ratio $R_b$, which we
interpret as a relative measurement of cross sections, and the $b$ quark
forward-backward asymmetry $A_{FB}(b)$ at three center of mass energies
on- and approximately $\pm 2$ GeV off-peak.  In addition, we
incorporated bounds from the DELPHI Collaboration on the ratio of $\sim
\pm 2$ GeV off-peak to on-peak values of $R_b$: $R_b^{-2}/R_b^0 = 0.982 
\pm 0.015$ and $R_b^{+2}/R_b^0 = 0.997 \pm 0.016$~\cite{DELPHI}.  These
constraints are stringent, as systematic errors cancel in the ratios.
Aside from the standard lineshape variables, namely, $m_Z$, $\Gamma_Z$,
$\Gamma(Z\to e^+e^-)$, and the hadronic peak cross section,
$\sigma^0_{\rm had}$, we simultaneously fit to the $\tilde{\nu}$ mass
and to its partial decay widths into $e^+e^-$ and $b\bar{b}$ pairs.

We find that a sneutrino near the $Z$ resonance is not excluded by the
high precision scans of the $Z$ lineshape.  After introducing the 3 fit
parameters associated with the sneutrino, the overall $\chi^2$ improves
significantly.  We find one minimum with $m_{\tilde{\nu}} = 91.79 \pm
0.54$ GeV, $\Gamma_{\tilde{\nu}} = 1.7^{+2.0}_{-1.4}$ GeV,
$\lambda_{131} = 0.013^{+0.004}_{-0.006}$, $\lambda^\prime_{333} =
0.56^{+ 0.27}_{- 0.30}$, and $\chi^2/\text{d.o.f.}=54.1/51$, relative to
60.6/54 in the SM.  The improvement in the fit comes primarily from
$R_b$ and $A_{FB}(b)$ as may be seen in Table~\ref{table:chi}.  The
sneutrino width is dominated by the partial decay width into $b\bar{b}$,
which in turn is strongly correlated with the width into $e^+e^-$
pairs. The reason for this, and for the large error range, is that the
$\tilde{\nu}$ peak cross section for $b\bar{b}$ pairs is, for a given
$m_{\tilde{\nu}}$, roughly determined by the $R_b$ data and given by
\begin{equation}
\sigma^0_{b\bar{b}} = \frac{8\pi 
\Gamma_{\tilde{\nu}\rightarrow e^+e^-}
\Gamma_{\tilde{\nu}\rightarrow b\bar{b}}}
{m_{\tilde{\nu}}^2 \Gamma_{\tilde{\nu}}^2} \approx 
\frac{8\pi \Gamma_{\tilde{\nu}\rightarrow e^+e^-}}
{m_{\tilde{\nu}}^2 \Gamma_{\tilde{\nu}\rightarrow b\bar{b}}} \ .
\end{equation}
The extracted $Z$ lineshape parameters are almost identical to the SM,
except that $\sigma^0_{\rm had}$ is reduced by 2/3 of a standard
deviation, slightly lowering the extracted $\alpha_s$.

\begin{table}
\caption{Total $\chi^2$/d.o.f. and $\chi^2$ contributions from $R_b$ 
and $A_{FB}(b)$ below-, on-, and above-peak, for our fit to the SM with
and without the $\tilde{\nu}$ resonance.}
\begin{tabular}{lccccccc}
& $\chi^2$/d.o.f. & $R_b^{-2}$ & $R_b^{0}$ & $R_b^{+2}$ & 
$A_{FB}^{-2}(b)$ & $A_{FB}^{0}(b)$ & $A_{FB}^{+2}(b)$ \\ \hline
SM               & 60.6/54  & 1.2 & 4.5 & 0.0 & 1.1 & 2.6 & 2.3 \\
SM+$\tilde{\nu}$ & 54.1/51  & 0.5 & 0.1 & 0.2 & 1.1 & 1.4 & 1.7 
\end{tabular}
\label{table:chi}
\end{table}

Fits with comparable $\chi^2$ also exist for $m_{\tilde{\nu}}$ below the
$Z$ peak.  In fact, an even better fit exists with $m_{\tilde{\nu}} =
90.28$ GeV, $\Gamma_{\tilde{\nu}} = 0.003$ GeV, $\lambda_{131} = 0.027$,
$\lambda'_{333} = 0.016$, and $\chi^2/\text{d.o.f.} = 53.6/51$.  This
narrow width minimum is made possible by initial state radiation, which
has the effect of broadening the $\tilde{\nu}$ resonance, allowing it to
improve the $\chi^2$ for scan points with $\sqrt{s} > m_{\tilde{\nu}}$.

It is important to note that the location of $m_{\tilde{\nu}}$ and the
size of the allowed window is largely dictated by the DELPHI off-peak
results for $R_b$. Omitting them would enlarge the window and also allow
an improvement in the prediction for $A_{FB}(b)$ at the peak+2 GeV
position.  We would like to encourage the other LEP groups to perform a
similar analysis of their off-peak data.

In conclusion, we have discussed the possibility of $\rlap{$\not$}R_P$
sneutrino resonances at LEP.  We find that the relevant
$\rlap{$\not$}R_P$ operators are only moderately constrained at present,
leaving open the possibility of a sneutrino width of several GeV.  Such
a resonance is the unique opportunity to directly probe 
supersymmetric masses up
to $\sqrt{s}$, greatly extending the reach in supersymmetry parameter
space, and could be discovered at LEP II either through analyses of
$\sqrt{s_{\text{eff}}}$ distributions or through additional low
luminosity runs at strategically chosen beam energies.  Furthermore, a
window with $m_{\tilde{\nu}} \approx m_Z$ exists and is currently
preferred by the data. We encourage 
the search for peaks in the $\sqrt{s_{\text{eff}}}$ distributions of the
forthcoming hadronic event samples, and the serious consideration of the
proposed additional low luminosity runs at LEP II energies.

\acknowledgments

It is pleasure to thank K.~Agashe, J.~Conway, A.~Falk, M.~Graesser,
H.~E.~Haber, L.~Hall, M.~Hildreth, M.~Suzuki, and especially T.~Han and
K.~M\"onig for useful discussions.  We are also grateful to the referee
for useful comments regarding ISR.  This work was supported in part by
the Director, Office of Energy Research, Office of High Energy and
Nuclear Physics, Division of High Energy Physics of the U.S. Department
of Energy under Contract DE--AC03--76SF00098, and in part by the NSF
under grants PHY--95--14797 and PHY--94--23002. JLF is a Research
Fellow, Miller Institute for Basic Research in Science and thanks the
high energy theory group at Rutgers University for its hospitality.


\begin{thebibliography}{99}

\bibitem{hs}
L.~J.~Hall and M.~Suzuki, Nucl. Phys. {\bf B231}, 419 (1984).

\bibitem{np} 
See {\em e.g.}, H.~P.~Nilles and N.~Polonsky, {\tt hep-ph/9606388},
Nucl. Phys. B, in press, and references therein.

\bibitem{dh88}
S.~Dimopoulos and L.~J. Hall, Phys.~Lett.~B {\bf 207}, 210 (1988);
S. Dimopoulos, R.~Esmailzadeh, L.~J.~Hall, J.-P.~Merlo, and
G.~D.~Starkman, Phys.~Rev.~D {\bf 41}, 2099 (1990); 
H.~Dreiner and S.~Lola, in {\em Proceedings of the Workshop on $e^+e^-$
collisions at 500 GeV}, 1991.

\bibitem{bgh} 
V.~Barger, G.~F. Giudice, and T.~Han, Phys.~Rev.~D {\bf 40}, 2987 (1989).

\bibitem{bes}
G.~Bhattacharyya, J.~Ellis, and K.~Sridhar, Mod. Phys. Lett.~A {\bf 10},
1699 (1995).

\bibitem{Btauexp}
F.~Behner, talk given at the EPS-HEP Conference, Brussels (1995).

\bibitem{Btauth}
A.~F.~Falk, Z.~Ligeti, M.~Neubert, and Y.~Nir, Phys. Lett.~B {\bf 326},
145 (1994).

\bibitem{hou}
W.-S.~Hou, Phys.~Rev.~D {\bf 48}, 2342 (1993).

\bibitem{PDG}
R.~M.~Barnett {\em et al.}, Phys.~Rev.~D {\bf 54}, 1 (1996).

\bibitem{ag} 
K.~Agashe and M.~Graesser, Phys.~Rev.~D {\bf 54}, 4445 (1996).

\bibitem{top} 
D.~Gerdes, {\tt hep-ex/9609013}.

\bibitem{qcd} 
See, for example, E.~L.~Berger and H.~Contopanagos, Phys.~Lett.~B {\bf
361}, 115 (1995).

\bibitem{Barger}
R.~M.~Godbole, P.~Roy, and X.~Tata, Nucl. Phys. {\bf B401}, 67 (1993);
V.~Barger, W.-Y.~Keung, and R.~J.~N.~Phillips, Phys.~Lett.~B {\bf 364},
27 (1995).


\bibitem{jens} 
J.~Erler, Phys.~Rev.~D {\bf 52}, 28 (1995).

\bibitem{ICHEP} 
A.~Blondel, plenary talk (Pl--9a) presented at the XXVIII International
Conference on High Energy Physics, July 1996, Warsaw, Poland.

\bibitem{L3} 
The L3 Collaboration, M.~Acciarri {\em et al.}, Z.~Phys.~C {\bf 62}, 551
(1994).

\bibitem{SLD} 
The SLD Collaboration, K.~Abe {\em et al.}, Phys. Rev. Lett. {\bf 70},
2515 (1993); {\em ibid.}, {\bf 73}, 25 (1994); E.~Torrence,
SLAC--PUB--7307, {\tt hep-ex/9610001}.

\bibitem{DELPHI} 
The DELPHI Collaboration, P.~Abreu {\em et al.}, Z.~Phys.~C {\bf 70},
531 (1996).

\end{thebibliography}
\end{document}